\documentclass[letterpaper,12pt]{article}
\pdfoutput=1 

\usepackage{jheppub}

\usepackage{tikz,tkz-graph,tkz-berge}
\usetikzlibrary{decorations.markings}
\usetikzlibrary{shapes.geometric}
\pgfdeclarelayer{edgelayer}
\pgfdeclarelayer{nodelayer}
\pgfsetlayers{edgelayer,nodelayer,main}
\tikzstyle{none}=[inner sep=0pt]
\tikzstyle{simple}=[-,draw=black,line width=1.000]

\usepackage{bbm}
\usepackage{amssymb,amsmath,amsthm}
\usepackage{graphicx,varwidth}
\usepackage[percent]{overpic}
\usepackage{mathtools}
\usepackage{xcolor}
\usepackage{natbib} 
\usepackage{physics}
\usepackage{soul}
\usepackage[colorlinks=true, allcolors=darkblue, urlcolor=darkblue, linktocpage=true, linkcolor=darkblue, citecolor=purple]{hyperref}

\colorlet{darkblue}{blue!70!black}
\colorlet{darkgreen}{green!70!black}

\newcommand{\nn}{\nonumber}

\newcommand\be{\begin{equation}}
\newcommand\ba{\begin{eqnarray}}
\newcommand\ee{\end{equation}}
\newcommand\ea{\end{eqnarray}}

\numberwithin{equation}{section}

\title{\boldmath Scrambling in Two-Dimensional Conformal Field Theories with Light and Smeared Operators}
\author[a]{Harsha R. Hampapura,}
\author[a,b]{Andrew Rolph,}
\author[a,c]{and Bogdan Stoica}

\affiliation[a]{Martin A. Fisher School of Physics,\\Brandeis University, Waltham, MA 02453, USA}
\affiliation[b]{Kavli Institute of Theoretical Physics, \\ University of Santa Barbara, Santa Barbara, CA 93106, USA}
\affiliation[c]{Department of Physics, Brown University, Providence RI 02912, USA}
\preprint{BRX-TH-6636, Brown-HET-1774}
\emailAdd{hrharsha@brandeis.edu}
\emailAdd{andrewrolph@brandeis.edu}
\emailAdd{bstoica@brandeis.edu}

\abstract{We study quantum chaos in two dimensional conformal field theories, building on the work analyzing the out-of-time order thermal correlation functions using large-$c$ Virasoro blocks. Our work investigates the contribution of light intermediate channels and smearing length scales to the four-point function and scrambling. Precise relations for how light intermediate channels increase the scrambling time and how smearing length scales smaller than the thermal length scale decrease the scrambling time are derived.}

\begin{document} 

\maketitle
\flushbottom

\section{Introduction}
Unitary quantum mechanical systems do not exhibit information loss in that they never forget their initial state, however observers without access to all degrees of freedom may not be able to distinguish microstates. If an observer with access to less than half the system's degrees of freedom cannot distinguish a perturbed state from an unperturbed one then the information contained in the perturbation is said to be scrambled. The time required for this dynamical chaotic mixing to occur is the scrambling time. 

In classical physics chaos is understood as sensitivity to initial conditions. If one considers the phase space coordinate $x(t)$ in a classical chaotic system and perturbs $x(0)$ an infinitesimal amount, one can diagnose the sensitivity to initial conditions through the Poisson bracket
\be
\{x(t),p(0)\}= \frac{\partial x(t)}{\partial x(0)}.
\ee
This quantity initially grows as a sum of exponentials in $t$, with the exponents called Lyapunov exponents. An analogous quantity to consider for a quantum system in state $\rho$ is the squared commutator~\cite{kitaevscrambling}
\be C(t) := - \Tr (\rho [W(t),V(0)]^2 ). \ee
At the onset of scrambling, $C(t)$ also grows exponentially with $t$, which from the perspective of operator growth arises from how the unitarily evolved $W(t)= e^{iHt}W(0)e^{-iHt}$ grows with time to become a larger sum of longer operator products, due to non-trivial commutations with the Hamiltonian. Loosely speaking, the commutator squared is determined by the fraction of operator products in $W(t)$ which contain $V(0)$. 

The systems we are especially interested in are many-body systems in thermal equilibrium, partly due to the AdS/CFT correspondence \cite{Maldacena:1997re} and the holographic dual description of black holes as thermofield double (TFD) states in double copies of large $c$ CFTs \cite{maldacena2003eternal}. For holographic field theories one can understand scrambling in thermal states as a disruption of the special TFD state entanglement between the left and right CFTs as diagnosed by mutual information between subregions in the two copies, which on the bulk side corresponds the lengthening of the wormhole connecting the two asymptotic regions~\cite{Shenker2013a,Shenker2013,leichenauer2014disrupting,Shenker2014}. The wormhole lengthens because low energy quanta produced far in the past become highly boosted near the black hole horizon, giving a large shockwave backreaction to the geometry.

Scrambling is also relevant to the black hole information paradox, where a remarkable result \cite{hayden2007black} shows that, in certain cases, information absorbed by a black hole can be emitted almost immediately after it has been scrambled amongst the black holes degrees of freedom. It has been conjectured \cite{Sekino2008,Susskind2011} that black holes scramble information faster than any other quantum system in nature, and some evidence for this conjecture has been found \cite{Lashkari2011}. 

Returning to our tool for studying quantum chaos, the squared commutator $C(t)$, expanding out gives four terms
\ba \label{OTOC}
C(t) &=&  \langle V(0)W(t)W(t)V(0) + W(t) V(0) V(0) W(t) \nn\\
& & - W(t) V(0) W(t) V(0)-V(0)W(t)V(0)W(t) \rangle_\beta.
 \ea 
This simplifies after the thermal relaxation time as $V$ acting on the thermal state becomes indistinguishable from the thermal state to local operators. The term $\langle V W W V \rangle_\beta$ can be understood as the expectation value of the operator $W W$ in the state obtained by $V$ acting on the thermal ensemble (and vice versa for $\langle WVVW \rangle_\beta$). If the energy inserted by the operator $V$ is small then after the dissipation time $\langle V W W V \rangle_\beta$ is given by the thermal expectation value $\langle W W \rangle_\beta$ multiplied by the norm of the state $\langle V V \rangle_\beta $. Thus, Eq. \eqref{OTOC} becomes,
\be \label{eq:last c}
C(t) = 2[\langle W(t) W(t) \rangle_\beta \langle V(0) V(0)\rangle_\beta - \Re\{\text{OTOC}(t) \}],
\ee
with the out of time ordered correlator (OTOC) defined by
\be OTOC(t) := \langle W(t) V(0) W(t) V(0) \rangle_\beta . \ee

OTOCs and $C(t)$ are thus equivalent ways of diagnosing chaos. At early times, the disconnected product and the OTOC terms in Eq. \eqref{eq:last c} cancel and the commutator squared is zero. At the onset of scrambling, the OTOC decays at a bounded rate $\lambda_L\leq 2\pi / \beta$~\cite{Maldacena2015,fitzpatrick2016quantum,perlmutter2016bounding}.
We define scrambling time as the operator time separation $t$ at which the OTOC is exponentially decaying to zero, or equivalently when $C(t)$ is asymptotically approaching the disconnected product $2\langle VV \rangle \langle WW \rangle$.

There is a significant body of literature on understanding scrambling from the holographic bulk perspective. In contrast, we will do a purely field-theoretic study, primarily building on work in~\cite{Stanford2015}, though we will give some holographic interpretation in the Discussion. The authors of \cite{Stanford2015} explicitly calculate the OTOC in two-dimensional conformal field theories using the known form of semiclassical Virasoro blocks \cite{Fitzpatrick2014}. Specifically they consider the contribution of the identity block to the OTOC, and from it, extract the scrambling time
\be \label{RS_scramblingtime}
t_* = \frac{\beta}{2\pi}\log \frac{c}{h_w} ,
\ee
with $h_w$ the holomorphic weight of the $W$ operator. Ideally, one would analyze scrambling for a pair of light operators with  both $h_v, h_w \ll c$, however the semiclassical conformal block is only valid for fixed $h_w /c$, corresponding to a heavy $W$ operator. In \cite{Roberts2014}, by matching with a bulk shockwave calculation, the authors conjecture the validity of the semiclassical formula in the light-light operator limit.

In this paper, we investigate the effect of non-identity Virasoro blocks and of smearing length scales on the scrambling time \eqref{RS_scramblingtime}. We examine the contribution of higher primaries and demonstrate that the scrambling time depends on the spectrum of the CFT,and that the existence of a light primary operator with $\mathcal{O}(1)$ OPE coefficients with the $V$ and $W$ operators, and conformal weights $h_p , \bar h_p \ll c$ bounds the scrambling time from below as
\be \label{eq:intro_light_st}
t_{*} \geq \frac{\beta}{2\pi}\left (\frac{2h_v + h_p}{\raisebox{-0.4ex}{$2h_v + \bar h_p$}}\right)\log \left(\frac{c}{h_w}\right),
\ee
with $h_v$ the holomorphic weight of the $V$ operator.  The scrambling time is determined by the primary operator for which the prefactor in \eqref{eq:intro_light_st} is the largest. Light operators with $h_p > \bar h_p$ increase the scrambling time. Note that primary operators with $\bar h_p > h_p$ do not violate the fast scrambling conjecture, as the existence of the identity operator bounds the scrambling time from below by \eqref{RS_scramblingtime}.

We also consider the scrambling of operator valued distributions, smearing the $V$ and $W$ operators over spatial scales $L_V, L_W$ in order to better understand the relation between the energy scale of perturbations and scrambling time. We calculate the scrambling time for smearing length scales much  smaller thermal length scale to be
\be
\label{eq:smeared scrambling time} 
t_* = \frac{\beta}{2\pi}\left(\log(\frac{c}{h_w}) - \log (\frac{\beta^2}{L_V L_W})\right),
\ee
corresponding  to a reduction in scrambling time due to high energy modes. For smearing length scales greater than $\beta$ we argue that the scrambling time increases without limit. The scrambling time \eqref{RS_scramblingtime} is the intermediate regime for perturbations with thermal scale energy.

The plan of this paper is as follows: In Section \ref{sec:background}, we set up the necessary CFT conventions and formulae to analyze the OTOC, in Section \ref{sec:light} we use the semiclassical Virasoro block to calculate the scrambling time in large $c$ two-dimensional CFTs with light higher-spin primaries. In Section \ref{sec:smeared} we calculate how the spatial smearing of the operators $V$ and $W$ changes the scrambling time, and in Section \ref{sec:discussion} we discuss the holographic interpretation of our results and future directions.

\textbf{Note:} This paper has some overlap in scope with work \cite{Liu2018}, as the two sets of authors worked in collaboration for much of the project until deciding to publish separately. A discussion of differences in analysis, results and interpretation is given in appendix \ref{appendix}.

\section{CFT background and conventions} \label{sec:background}
We are interested in the thermal four-point correlation function involving two operators $V$ and $W$ separated by Lorentzian time $t$ and spatial distance $x$. Correlation functions in the thermal state can be mapped to expectation values in the vacuum using the conformal transformation
\be
z(x,t) = e^{\frac{2\pi}{\beta}(x+t)}, \qquad \bar z (x,t) = e^{\frac{2\pi}{\beta}(x-t)},
\ee
with correlators of primary operators related by
\be
\langle \mathcal{O}_{h,\bar h} (x,t) ... \rangle_\beta = \left (\frac{2\pi z}{\beta}\right)^h \left(\frac{2\pi\bar z}{\beta}\right)^{\bar h} \langle \mathcal{O}_{h,\bar h} (z,\bar z)... \rangle.
\ee

The specific correlation function we wish to work with is a four point function, normalised by the product of two-point functions,
\be \label{eq:ratio}
\frac{\langle W (z_1,\bar z_1)W (z_2,\bar z_2)V (z_3,\bar z_3)V (z_4,\bar z_4)\rangle}{\langle W (z_1,\bar z_1)W (z_2,\bar z_2)\rangle \langle V (z_3,\bar z_3)V (z_4,\bar z_4)\rangle}.
\ee
Following the canonical choice, we use the global conformal transformations SL(2,$\mathbb{C}$) to take $z_1, \bar z_1 \to \infty; z_2 , \bar z_2 \to 1; z_4 , \bar z_4 \to 0$ \cite{DiFrancesco:1997nk}. With this choice, the holomorphic cross ratio is
\be
z:= \frac{z_{12} z_{34}}{z_{13}z_{24}} = z_3,
\ee
and similarly $\bar z = \bar z_3$, and the ratio \eqref{eq:ratio} becomes
\be \begin{aligned} \label{eq:ratio2}
\frac{\langle W (z_1,\bar z_1)W (z_2,\bar z_2)V (z_3,\bar z_3)V (z_4,\bar z_4)\rangle}{\langle W (z_1,\bar z_1)W (z_2,\bar z_2)\rangle \langle V (z_3,\bar z_3)V (z_4,\bar z_4)} &= \lim_{z_1 , \bar z_1 \to \infty} z_1^{2h_w}\bar z_1^{2 \bar h_w} z^{2h_v}\bar z^{2 \bar h_v} \langle W (z_1,\bar z_1)W (1,1)V (z,\bar z)V (0,0)\rangle \\
&= \sum_p C_{VV}^p C_{WW}^p z^{2h_v} \mathcal{F}_p(z) \bar z^{2\bar h_v} \mathcal{F}_p(\bar z),
\end{aligned} \ee
where, in the second line, we have expanded in the $z \rightarrow 0$ channel to write the four-point function in terms of the Virasoro conformal blocks. From Eq. \eqref{eq:ratio2} we see that the quantity of interest when comparing the OTOC to the disconnected product is $z^{2h_v} \mathcal{F}_p(z)$ and its anti-holomorphic counterpart.\footnote{The factor of $z^{2h_v}$ agrees with the convention of \cite{Roberts2014} and is non-standard.}

\section{Light intermediate channels} \label{sec:light}
In this section we investigate the importance of intermediate channels to the OTOC, their dominance over the identity block and subsequent effect on the scrambling time. In putative bulk theories, this corresponds to the exchange of massive particles. Following \cite{Stanford2015}, we will work in the semiclassical limit $c \gg 1$ where the conformal blocks $\mathcal{F}_p (z)$ exponentiate,

\begin{equation}
\mathcal{F}_p (z) = \exp (- \frac{c}{6}f_p (z)),
\end{equation}
and use a result from \cite{Fitzpatrick2014} for the semiclassical conformal block, valid for $c \gg 1$; $h_v, h_p \ll c$; $h_w/c$ fixed but arbitrary,
\begin{align}
\frac{c}{6}f(z) & =(2h_{v}-h_{p})\log\left(\frac{1-(1-z)^{\alpha_{w}}}{\alpha_{w}}\right)+h_{v}(1-\alpha_{w})\log(1-z) \nn \\
 & \qquad+2h_{p}\log\left(\frac{1+(1-z)^{\alpha_{w}/2}}{2}\right),
\end{align}
with $\alpha_w := \sqrt{1-24h_w/c}$. This gives a conformal block
\begin{equation} \label{eq:conformal block principal sheet}
z^{2h_v} \mathcal{F}_{p}(z)=\left[\frac{\alpha_{w}z(1-z)^{(\alpha_{w}-1)/2}}{1-(1-z)^{\alpha_{w}}}\right]^{2h_{v}}\cdot\left[\frac{4}{\alpha_{w}}\frac{1-(1-z)^{\alpha_{w}/2}}{1+(1-z)^{\alpha_{w}/2}}\right]^{h_{p}}.
\end{equation} 

Following the procedure detailed in \cite{Stanford2015} we consider the analytic continuation of the Euclidean four-point function to Lorentzian time which gives the OTOC's ordering of operators. This analytic continuation causes $z$ to pass around its branch point at $z=1$ and hence for $z^{2h_v} \mathcal{F}_p (z)$ to pass to the second Riemann sheet. $\bar z$ does not pass around its branch point at $\bar z = 1$. The conformal block on the second sheet is
\begin{equation}
z^{2h_v} \mathcal{F}_{p}^{II}(z)=\left[\frac{\alpha_{w}ze^{-\pi(\alpha_{w}-1)i}(1-z)^{(\alpha_{w}-1)/2}}{1-e^{-2\pi i\alpha_{w}}(1-z)^{\alpha_{w}}}\right]^{2h_{v}}\cdot\left[\frac{4}{\alpha_{w}}\frac{1-e^{-\pi\alpha_{w}i}(1-z)^{\alpha_{w}/2}}{1+e^{-\pi\alpha_{w}i}(1-z)^{\alpha_{w}/2}}\right]^{h_{p}}.
\end{equation}

\begin{figure} 
\begin{overpic}[width=0.7\paperwidth]{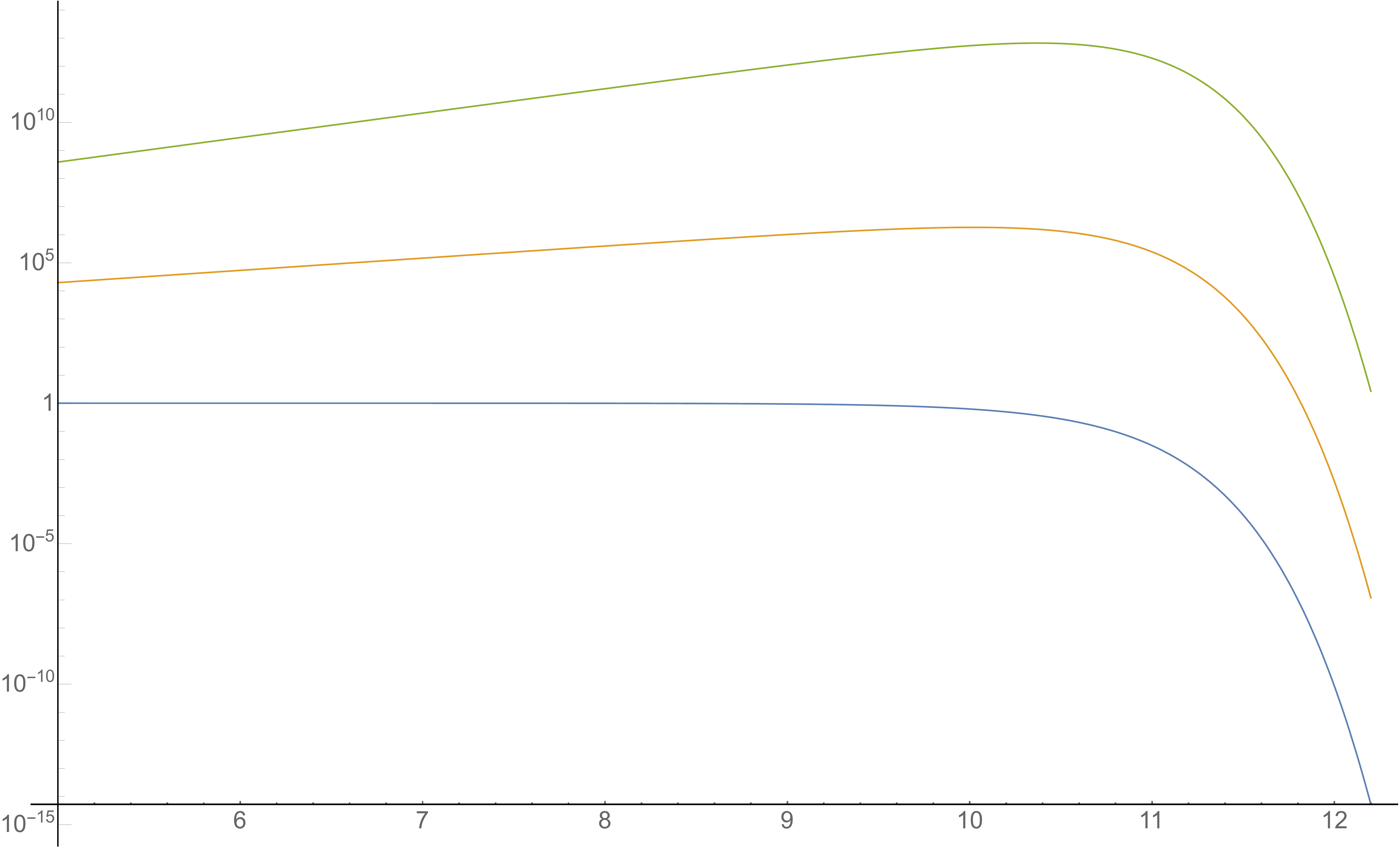}
\put(0,62){\colorbox{white}{\parbox{0.1\linewidth}{%
     $|z^{2h_v}\mathcal{F}^{II}_p (z)|$
     }}}
\put(100,2){\colorbox{white}{\parbox{0.1\linewidth}{%
     $t$
     }}}
     \put(42,34){\colorbox{white}{\parbox{0.4\linewidth}{%
     $h_p = 0$ (Identity block)
     }}}
     \put(68,46){\colorbox{white}{\parbox{0.1\linewidth}{%
     $h_p =1$
     }}}
     \put(72,59){\colorbox{white}{\parbox{0.1\linewidth}{%
     $h_p =2$
     }}}

\end{overpic}
\caption{Conformal block $|z^{2h_v} \mathcal{F}_{p}^{II}(z)|$ against time separation $t$ for different values of intermediate channel holomorphic weight $h_{p}$, for $h_{w}=5,h_{v}=1000$ and $c=10^{9}$. Shows the initial exponential growth with $t$ of the conformal block with $h_p > 0$, the decay around scrambling time, and the dominance of light intermediate conformal blocks over the identity block.}\label{fig:light weight channels}
\end{figure}
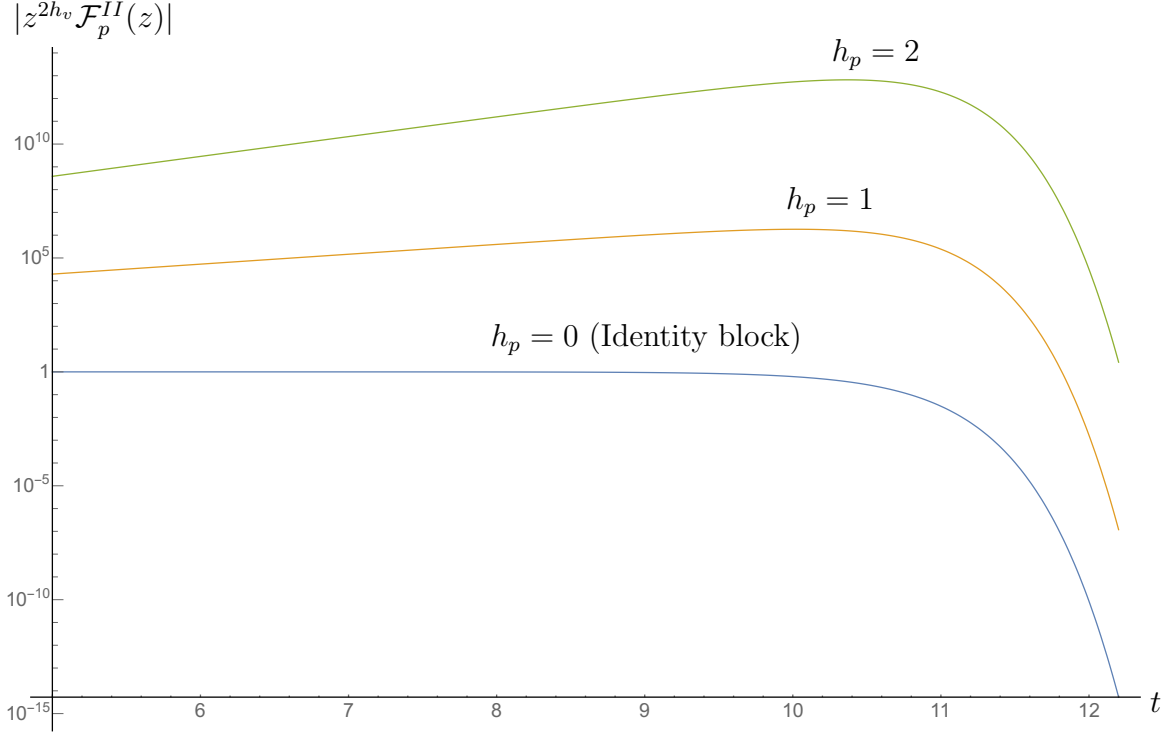

We are interested in the behaviour of the conformal block in three different time regimes: before the fast scrambling time~\eqref{RS_scramblingtime} but after the dissipation time $t \sim \beta$, around the fast scrambling time, and after the fast scrambling time, corresponding to $ h_w /c \ll z \ll 1$, $ z \sim h_w /c$, and $z \ll h_w /c$ respectively. In these three regimes the conformal block takes the form 
\begin{equation} \label{eq:holomorphic conformal block}
    z^{2h_v} \mathcal{F}_{p}^{II}(z)\approx 
\begin{dcases}
    \left(\frac{{16}}{z}\right)^{h_{p}} &\qquad\ h_w /c \ll z \ll 1 \\
    \left(\frac{16}{z}\right)^{h_{p}}\left(\frac{1}{1-\frac{24\pi i h_w}{c z}}\right)^{2h_v+h_p}            &\qquad\qquad z \sim h_w /c \\
16^{h_p}\left(\frac{ic}{24\pi h_w}\right)^{2h_v+h_p} z^{2h_v}              &\qquad\qquad z \ll h_w /c   
\end{dcases}.
\end{equation}
To see how this function depends on $t$ recall that $z \sim e^{-\frac{2\pi}{\beta} t}$. Figure~\ref{fig:light weight channels} illustrates how the conformal block depends on the operator time separation for a few values of $h_p$. For the identity block with $h_{p}=0$ it is constant at early times when $z \gg h_w /c$, then starts to exponentially decay around $z \sim h_w /c$. The light intermediate channels have significantly different behaviour, they initially grow exponentially with $t$ and then start to decay roughly around the same time as the identity block. 

For the OTOC, one also needs to consider the effect of the antiholomorphic conformal block factor $ \bar z^{2\bar h_v} \mathcal{F}_p(\bar z)$. The key difference from the holomorphic block is that during the analytic continuation of the Euclidean four point function, the holomorphic cross ratio $z$ passes around its branch point at $z = 1$ and goes to the second Riemann sheet, while the antiholomorphic cross ratio does not and stays on the principal sheet. The principal sheet antiholomorphic block acts to suppress the exponential growth of the second sheet holomorphic block,as seen by replacing holomorphic variables in~\eqref{eq:conformal block principal sheet} with their antiholomorphic counterparts, then taking $\bar z \ll 1$,
\be
 \bar z^{2 \bar h_v}\mathcal{F}_p(\bar z) \approx (2 \bar z)^{\bar h_p}  \propto \exp (- \frac{2\pi}{\beta} \bar h_p (t+ x)).
\ee
For the identity block with $\bar h_p = 0$ this factor is trivial and irrelevant to the scrambling time derivation in~\cite{Roberts2014}. However, for $\bar h_p >0$, this factor acts to suppress the contribution of light primary operators, it is an exponentially decaying function in $t$. We will discuss the case $\bar h_p < 0$ after combining this antiholomorphic conformal block with its second sheet holomorphic counterpart, to find the behaviour of the OTOC with $t$.

At early times, though with $t \gg \beta$, the product of conformal blocks grows exponentially with $t$ as 
\be \label{eq:growing with t}
z^{2h_v} \mathcal{F}_{p}^{II}(z)  \bar z^{2 \bar h_v}\mathcal{F}_p(\bar z) \propto \exp (\frac{2\pi}{\beta}(h_p - \bar h_p) t),
\ee and surprisingly this contribution to the OTOC dominates over the identity block at $t$ given by \eqref{RS_scramblingtime}, assuming $h_p > \bar h_p$ and generic OPE coefficients. Around this time the second expression in \eqref{eq:holomorphic conformal block} shows that all the holomorphic conformal blocks switch from exponential growth in $t$ to exponential decay. The exception is the identity block as $|z^{2h_v} \mathcal{F}_{0}^{II}(z)|$ is a monotonically decreasing function in time, starting at $1$ before starting its exponential decay as scrambling takes effect. The holomorphic conformal blocks for the other primary operators reach their maximum magnitude at
\begin{equation} \label{eq:zmax}
z_{max}(h_p)=2 \pi(1-\alpha_{w})\sqrt{\frac{2h_{v}}{h_{p}}},
\end{equation}
corresponding to a time separation
\begin{equation}
t_{max} =\frac{\beta}{2\pi}\log (\sqrt{\frac{h_p}{h_v}}\frac{ c}{h_w }).
\end{equation}

At late times, all holomorphic blocks decay at the same rate determined by the third equation in \eqref{eq:holomorphic conformal block}, $\exp (-\frac{4\pi}{\beta}h_v t)$,  and so the light intermediate states dominate over the identity block for all times with no crossover. Including the contribution of the antiholomorphic block gives
\be
z^{2h_v} \mathcal{F}_{p}^{II}(z)  \bar z^{2 \bar h_v}\mathcal{F}_p(\bar z) \sim \left(\frac{c}{h_w}\right)^{2h_v + h_p}\exp (-(2h_v + \bar h_p)\frac{2\pi}{\beta}t)
\ee
at late times, when $z \ll h_w /c$. What we are most interested in is the value of $t$ for which the commutator squared $C(t)$ approaches $2\langle W(t) W(t) \rangle_\beta \langle V(0) V(0) \rangle_\beta$, or equivalently when the $|$OTOC $|$$\ll 1$. The time taken for $z^{2h_v}\mathcal{F}_p^{II} (z) \bar z^{2\bar h_v}\mathcal{F}_p (\bar z)$ to decay to an $\mathcal{O}(1)$ value is
\be \label{eq:lst}
t_{*p} = \frac{\beta}{2\pi}\left (\frac{2h_v + h_p}{\raisebox{-0.4ex}{$2h_v + \bar h_p$}}\right)\log \left(\frac{c}{h_w}\right).
\ee
Assuming OPE coefficients that are not parametrically small in powers of $h_w /c$ in order to suppress the contribution of light primary operators, the scrambling time is determined by the primary operator for which the prefactor in \eqref{eq:lst} is largest, as the OTOC is dominated by the conformal block for that operator. Roughly speaking the larger the spin the longer the scrambling time. As explained in the Introduction, the existence of primary operators with $\bar h_p > h_p$ does not lead to a shorter scrambling time, the scrambling time in the 2D CFTs we are considering is bounded from below by \eqref{RS_scramblingtime}. 

The apparent asymmetry between $h_p$ and $\bar h_p$ occurs because we are looking at Lorentzian correlators and this breaks the symmetry between holomorphic and antiholomorphic sectors of the CFT. In parity invariant CFTs, where for each primary operator with $(h_p , \bar h_p)$ there is a conjugate operator with weights $(\bar h_p , h_p)$, the scrambling time is parity invariant. While the decay time for a given block, given by \eqref{eq:lst}, is not invariant under $h_p \leftrightarrow \bar h_p$, the OTOC from which the scrambling time is derived is a sum over the full spectrum, which will be parity invariant. Assuming the parity invariant CFT has light operators and generic OPE coefficients, the scrambling time is increased.\footnote{We thank Albion Lawrence and Eric Perlmutter for this point.}

\section{Spatially smeared operators}
\label{sec:smeared}

 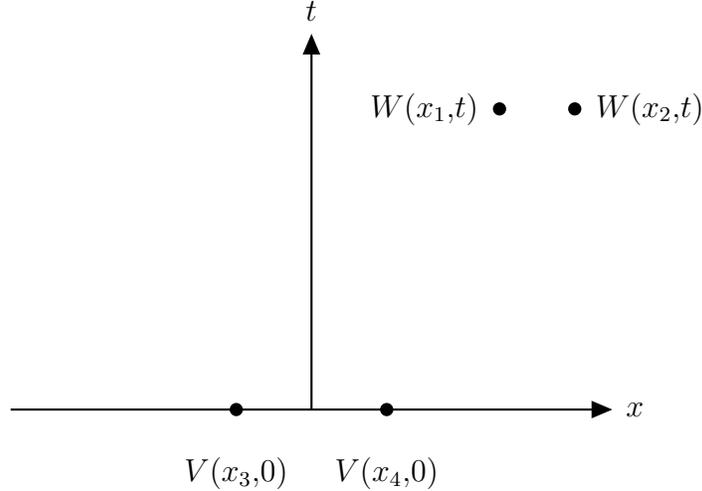
\begin{figure} 
 \centering 
 \begin{tikzpicture}
    
    \tikzset{VertexStyle/.style = {shape = circle,fill = white, minimum size = 0pt,inner sep = 0pt}}
    \Vertex[L=$ $,x=-4,y=0]{v0}
    \Vertex[L=$ $,x=4,y=0]{v1}
    \Vertex[L=$ $,x=0,y=0]{v2}
    \Vertex[L=$ $,x=0,y=5]{v7}
    
    \tikzset{VertexStyle/.style = {shape = circle,fill = black, minimum size = 5pt,inner sep = 0pt}}
    
    \Vertex[L=$ $,x=-1,y=0]{v3}
    \Vertex[L=$ $,x=1,y=0]{v4}
    \Vertex[L=$ $,x=2.5,y=4]{v5}
    \Vertex[L=$ $,x=3.5,y=4]{v6}
    
    \tikzset{VertexStyle/.style = {shape = circle,fill = white, minimum size = 0pt,inner sep = 0pt}}
    
    \EA[unit=0.3,L=$x$](v1){l1}
    \NO[unit=0.3,L=$t$](v7){l2}
    
    \SO[unit=0.85,L=$V(x_3{,}0)$](v3){l3}
    \SO[unit=0.85,L=$V(x_4{,}0)$](v4){l4}
    \WE[unit=1,L=$W(x_1{,}t)$](v5){l5}
    \EA[unit=1,L=$W(x_2{,}t)$](v6){l6}    
    
    \Edge[style={->,>=triangle 45}](v0)(v1)
    \Edge[style={->,>=triangle 45}](v2)(v7)
    
\end{tikzpicture}
 \caption{Positions of the point $V$ and $W$ operators prior to integration against Gaussian smearing functions of width $L_V$ and $L_W$.}
 \label{fig:operators}
 \end{figure}
In this section we will change gears and consider a different computation, in which we smear the operators over finite length scales. We would like to understand the effect of these scales on the scrambling time, to see how scrambling depends on the perturbations' energies. Let us introduce our set up. The first step in the smearing procedure is to consider point operators that are not spatially coincident and then  integrate against smearing functions. After the smearing procedure, each operator in a given pair will have finite spatial support and be centered about the same spatial position. Our labelling of the four operator positions is shown in Figure~\ref{fig:operators}. After the conformal mapping given by
\begin{equation}
\begin{aligned}z_{1} & =e^{\frac{2\pi}{\beta}(x_{1}+t)}, \qquad  \bar z_{1} =e^{\frac{2\pi}{\beta}(x_{1}-t)}\\
z_{2} & =e^{\frac{2\pi}{\beta}(x_{2}+t)},  \qquad \bar z_{2}  = e^{\frac{2\pi}{\beta}(x_{2}-t)}\\
z_{3} & =e^{\frac{2\pi}{\beta}x_{3}}, \qquad \quad \bar z_{3}  =e^{\frac{2\pi}{\beta}x_{3}}\\
z_{4} & =e^{\frac{2\pi}{\beta}x_{4}},\qquad \quad \bar z_{4}  =e^{\frac{2\pi}{\beta}x_{4}}
\end{aligned}
\end{equation}
the holomorphic and antiholomorphic cross-ratios
$z=z_{12}z_{34}/z_{13}z_{24}$ and $\bar z=\bar z_{12} \bar z_{34}/ \bar z_{13} \bar z_{24}$ are
\ba
z&=&\frac{\sinh(\frac{\pi}{\beta}(x_{1}-x_{2}))\sinh (\frac{\pi}{\beta}(x_{3}-x_{4}))}{\sinh(\frac{\pi}{\beta}(t+x_{1}-x_{3}))\sinh(\frac{\pi}{\beta}(t+x_{2}-x_{4}))}, \\
\bar z&=&\frac{\sinh(\frac{\pi}{\beta}(x_{1}-x_{2}))\sinh (\frac{\pi}{\beta}(x_{3}-x_{4}))}{\sinh(\frac{\pi}{\beta}(t-x_{1}+x_{3}))\sinh(\frac{\pi}{\beta}(t-x_{2}+x_{4}))}.
\ea
We take $t$ to be much larger than any $x_{i}$, otherwise the $V$ and $W$ operators have support on spacelike separated regions and by causality those components of the operators commute. In this limit the cross ratio becomes 
\begin{equation}
\label{eq:simp z}
z = 4 \sinh(\frac{\pi}{\beta}(x_{1}-x_{2}))\sinh (\frac{\pi}{\beta}(x_{3}-x_{4})) e^{-\frac{2\pi}{\beta}t}e^{\frac{\pi}{\beta}(-x_1 - x_2 + x_3 + x_4)}.
\end{equation}
From \eqref{eq:holomorphic conformal block} we recall that the second sheet identity block is 
\begin{equation} \label{eq:RS} z^{2h_v} \mathcal{F}^{II}(z) \approx \left ( \frac{1}{1-\frac{24\pi i h_w}{cz}} \right )^{2h_v}.
\end{equation}
This has a dependence on the four operator spatial positions $x_i$, through the dependence of $z$ on $x_i$, this is what we integrate against our smearing functions. We integrate the point operators $V$ and $W$ against Gaussians smearing functions with spatial width $L_V$ and $L_W$. The $V$ operator will be centred around $x = 0$ while the $W$ has a spatial offset, it is smeared about $x = x_W$, defined with the appropriate prefactor,
\begin{equation}
W_{smear} (t,x_W) := \frac{1}{L_W \sqrt{\pi}} \int_{-\infty}^{\infty} e^{-(x-x_W)^2 / L_W^2} W(t,x) dx.
\end{equation}
This is an operator at Lorentzian time $t$, smeared over a length scale $L_W$ about a central position $x_W$. The conformal block for the four smeared operators is then
\begin{equation} \label{eq:integral}
\int_{-\infty}^{\infty} dx_1 dx_2 dx_3 dx_4\frac{1}{L_W^2 L_V^2 \pi^2} \exp (-\frac{x_1^2}{L_V^2}- \frac{x_2^2}{ L_V^2} -\frac{(x_3-x_W)^2}{ L_W^2}-\frac{(x_4-x_W)^2}{ L_W^2}) z^{2h_v} \mathcal{F}(z).
\end{equation}
In terms of the integral, smearing length scale affects scrambling time because the parts of the integration region which contribute significantly depend on $L_V$ and $L_W$. Similar to the unsmeared identity block, we consider the perturbation to have been scrambled once the conformal block switches from being constant to exponential decaying with $t$. The non-Gaussian part of the integrand, given by \eqref{eq:RS}, has two asymptotic $z$ regimes: for early times with $z \gg h_w /c$ it is $1$, while for late times with $z \ll h_w /c$ it is
\begin{equation}\label{eq:small z}
  z^{2h_v} \mathcal{F}(z) \approx  \left(\frac{cz}{24\pi i h_w} \right )^{2h_v} \ll 1.
\end{equation}
Note that from \eqref{eq:simp z} $z$ is small for late times, or when the separation of operators within a pair is much less than the thermal length scale. The smaller the smearing length scale, the more the Gaussian smearing functions suppress large $x_i$, and so the dominant contribution to the integral is for $z \ll h_w /c$ with the conformal block given by \eqref{eq:small z}. For $L_V, L_W \ll \beta$ separations of operators within a pair of order the thermal length scale and larger are suppressed, and we can approximate the cross ratio \eqref{eq:simp z} by
\be \label{eq:z for small L}
 z \approx  \frac{4\pi^2}{\beta^2} (x_{1}-x_{2})(x_{3}-x_{4}) e^{-\frac{2\pi}{\beta}(t-\frac{x_3 + x_4}{2})}.
\ee
Combining \eqref{eq:z for small L} and \eqref{eq:small z} we can exactly evaluate the integral \eqref{eq:integral}, giving the smeared conformal block
\be \label{eq:done integral}
\frac{1}{4\pi}\left((1+e^{2\pi i h_v}) \Gamma\left(\frac{1}{2} + h_v\right)\right)^2\left( \frac{\pi c}{3 i h_w}\frac{L_V L_W}{\beta^2}  e^{-\frac{2\pi }{\beta}(t-x_W)}\right)^{2h_v},
\ee
Note that if $h_v$ is half-integer then this expression is exactly zero, however in discussing scrambling we wish to consider generic operators $V$, for which $h_v$ will not be half-integer. The time separation at which \eqref{eq:done integral} becomes $\mathcal{O}(1)$, marking the scrambling time, is 
\be \label{eq:smeared scrambling time} t_* = \frac{\beta}{2\pi}\left(\log(\frac{c}{h_w}) - \log (\frac{\beta^2}{L_V L_W})\right), \ee
valid for $L_V, L_W \ll \beta$. The smaller the smearing length scale, the larger the energy scale of the operator and the faster the perturbation is scrambled. When $L_V$ and $L_W$ are sufficiently small they can decrease the scrambling time at leading order, however as these high energy perturbations are not a small perturbation to the thermal state this does not violate the fast scrambling conjecture.

For operators smeared over length scales much larger than the thermal length scale, the region of $\{x_i\} \in \mathbbm{R}^4$ for which the Gaussians in \eqref{eq:integral} give support grows beyond the region of size $\beta^4$ centered around $(x_1, x_2 , x_3, x_4) = (0,0,x_W,x_W)$, and the $z \gg h_w /c $ limit for which the conformal block is $1$ becomes the important $z$ limit. The larger one takes $L_V$ and $L_W$, the larger the region in $\mathbbm{R}^4$ which is both not suppressed by the Gaussians and has $z\gg h_w /c$, and so the closer the smeared conformal block is to one for fixed $t$. By increasing the smearing length scale one increases the scrambling time. This is valid until one one reaches smearing length scales as wide as the lightcone. Interpolating between the result \eqref{eq:smeared scrambling time} for $L_V, L_W \ll \beta$ and the $L_V, L_W \gg \beta$ behaviour just described  to $L_V, L_W \sim \beta$ one concludes that operators smeared over the thermal length scale exhibit fast scrambling, at least when considering only the contribution of the identity block.

\section{Discussion}\label{sec:discussion}
In this section we discuss the holographic interpretation of our results, the relation to the chaos bound, and possible future work.

Let us first give a holographic interpretation for the dependence of scrambling time on smearing length scale. Perturbing the thermal state of a holographic 2D CFT with a single trace operator smeared over spatial length $L$ is dual to releasing a particle of energy $E \sim L^{-1}$ from the asymptotic boundary of a BTZ black hole. As we increase the smearing length scale, we reduce the energy of the bulk particle. Near the horizon, time translation corresponds to a boost, such that on the $t = 0$ slice a particle of energy $E$ released at the boundary at time $t = -t_w$ has proper energy
\be \label{eq:boost} E_p \sim \frac{E\beta}{2\pi l_{AdS}} \exp (\frac{\beta}{2\pi}t_w). \ee
Scrambling occurs when $t_w$ is large enough that the proper energy becomes of order $G_N^{-1} \sim c$, then the backreaction on the BTZ geometry becomes significant, leading to the lengthening of the wormhole and destruction of entanglement between the two CFT copies. If we increase $L$ or equivalently decrease $E$ then it takes a longer time to scramble. Equation \eqref{eq:boost} suggests that the dependence of scrambling time on the smearing length of the $W$ operator is $(\beta / 2\pi)\log (\beta / L_W)$, this is consistent with our result \eqref{eq:smeared scrambling time}.

The dominating contribution of higher dimension and spin primaries to the scrambling time in the CFT corresponds to bulk to bulk propagation of massive spin fields between the $V$ and $W$ fields. Assuming that the two-dimensional CFTs we have been studying that seem to have large scrambling times do in fact exist, and that they have a semiclassical quantum gravity dual, it is puzzling from the bulk perspective why the massive fields dual to the light intermediaries should increase the scrambling time.

The arbitrarily rapid exponential growth of the conformal block seen in Eq. \eqref{eq:growing with t} for $\beta \ll t \ll (\beta/2\pi) \log c$ does not violate the chaos bound. The chaos bound on Lyapunov exponents is derived by showing that functions $f(t)$ that are analytic and bounded $|f| \leq 1$ on the half-strip given by $\Re{t}>0$, $|\Im{t}|\leq \beta /4$ satisfy the inequality
\be \frac{1}{1-|f|}\left|\frac{df}{dt}\right| \leq \frac{2\pi}{\beta}+ \mathcal{O}\left(\exp (-\frac{4\pi}{\beta}t)\right), \ee
then assuming the ratio of the OTOC to the disconnected product is of the form
\be
\frac{\langle V W(t) V W(t) \rangle_\beta}{\langle VV\rangle_\beta \langle W(t) W(t) \rangle_\beta} \approx 1 - \epsilon e^{\lambda_L t} + ...
\ee
satisfy the above requirements for $f$, and one finds $\lambda_L \leq 2\pi /\beta$. The conformal block for light primaries we study in Section \ref{sec:light} is not of this form and does not obey the assumptions made in deriving the chaos bound. It is unusual in that it grows rather than decays with $t$, at an arbitrarily fast rate and to a value exponentially larger than the disconnected product. The chaos bound does not constrain the growth of the OTOC, so there is no inconsistency. That said, the conformal block is still a bounded function in $t$ and assuming it is analytic and bounded on the whole half strip its rate of decay is bounded after reaching its maximum. 

The early growth of the OTOC seems strange, as one usually interprets the decay of the OTOC as the onset of chaos, growth seems to imply increasing order in a chaotic system. Moreover, there is an argument that in large $N$ systems one expects the OTOC to be parametrically close to the disconnected product for all $t \gg \beta$, as detailed in section 4.3.1 of \cite{Maldacena2015}. In contrast, we found that each conformal block with $h_p > \bar h_p$ is exponentially larger than the disconnected product until long after the fast scrambling time. We are unclear as to the resolution of this apparent tension. It would be interesting to investigate if including heavy primaries makes the sum over Virasoro blocks divergent, and so necessitates a resummation.

In Section \ref{sec:light} we considered only light primaries with $\Delta \ll c$, but it is interesting to consider what effect primary operators of dimension $\Delta \sim c$ and heavier have on the OTOC and scrambling time. This question can partially be answered using the known semiclassical conformal blocks formula in the limit $h_p \to \infty$ with $c/h_p$, $h_w/c$ and $h_v/c$ all small and fixed~\cite{zamolodchikov1987conformal}, however we are not aware of formulae for intermediate weight intermediates $h_p \sim c$.

A natural extension of our analysis is the study of the OTOC in higher dimensional CFTs. Although some results are available for a class of interesting models \cite{Swingle2017}, a general analysis of the Virasoro blocks is rendered difficult by the absence of semiclassical results for the conformal blocks (which can be attributed to the presence of only global descendants). We leave this interesting question for future work.

\acknowledgments

AR, HRH, and BS would like to thank Matthew Headrick, Albion Lawrence, Eric Perlmutter, and Douglas Stanford for useful feedback on early versions of this draft, and Chang Liu and David Lowe for early collaboration. This work was supported in part by the U.S. Department of Energy under grant DE-SC0009987, and by the Simons Foundation through the It from Qubit Simons Collaboration on Quantum Fields, Gravity and Information.  AR is also supported by a KITP Graduate Fellowship, and is thankful to the Kavli Institute of Theoretical Physics for hospitality, and to Adam Levine for useful discussions. HRH would like to thank Pawel Caputa, Sunny Guha, Sunil Mukhi and Pranjal Nayak for useful discussions. BS thanks Jennifer Lin for discussions.

\appendix

\section{Remarks on reference \cite{Liu2018}}
\label{appendix}
Here we note some differences in our paper compared to \cite{Liu2018}. Our papers overlap in scope in that they study the effect of smeared operators and higher primaries on scrambling time. Let us highlight some differences in analysis and results to aid readers wishing to understand the results in the two papers. 

\begin{itemize}

\item With regards the light primaries, our scrambling time $t_{*p}$ is not related to the $t_{*p}$ given in equation (12) of \cite{Liu2018}. Our $t_{*p}$, for the dominating primary operator, corresponds to the time at which the commutator squared will approach the disconnected product $2 \langle VV \rangle \langle W(t) W(t) \rangle$. In \cite{Liu2018} the $t_{*p}$ computed is the time at which the ratio of the light primary operator block to the identity block saturates to a constant, which is not directly related to the scrambling time. The authors of \cite{Liu2018} also did not consider the contribution of the antiholomorphic conformal block to the OTOC.

\item Our smearing analysis also differs from that in \cite{Liu2018}. We smear each operator over its own Gaussian wavepacket. Reference \cite{Liu2018} smears one $V$ operator over a half-Gaussian on the positive $x$-axis, one over a mirror image half-Gaussian on the negative $x$-axis, and similarly for the $W$ operators. Varying $L$ then convolutes two separate effects on scrambling time, operator smearing length scale and spatial offset. 
\end{itemize}

\bibliographystyle{JHEP-custom}
\bibliography{scrambling}

\providecommand{\href}[2]{#2}\begingroup\raggedright\begin{thebibliography}{10}

\bibitem{kitaevscrambling}
A.~Kitaev, ``Hidden correlators in the hawking radiation and thermal noise.''
  Talk given at the Fundamental Physics Prize Symposium, 2014.

\bibitem{Maldacena:1997re}
J.~M. Maldacena, {\it {The Large N limit of superconformal field theories and
  supergravity}},  Int. J. Theor. Phys. {\bf 38} (1999) 1113--1133,
  \href{http://arxiv.org/abs/hep-th/9711200}{{\tt hep-th/9711200}}. [Adv.
  Theor. Math. Phys.2,231(1998)].

\bibitem{maldacena2003eternal}
J.~Maldacena, {\it Eternal black holes in anti-de sitter},  Journal of High
  Energy Physics {\bf 2003} (2003), no.~04 021.

\bibitem{Shenker2013a}
S.~H. Shenker and D.~Stanford, {\it {Multiple Shocks}},  JHEP {\bf 12} (2014)
  046, \href{http://arxiv.org/abs/1312.3296}{{\tt arXiv:1312.3296}}.

\bibitem{Shenker2013}
S.~H. Shenker and D.~Stanford, {\it {Black holes and the butterfly effect}},
  JHEP {\bf 03} (2014) 067, \href{http://arxiv.org/abs/1306.0622}{{\tt
  arXiv:1306.0622}}.

\bibitem{leichenauer2014disrupting}
S.~Leichenauer, {\it Disrupting entanglement of black holes},  Physical Review
  D {\bf 90} (2014), no.~4 046009.

\bibitem{Shenker2014}
S.~H. Shenker and D.~Stanford, {\it {Stringy effects in scrambling}},  JHEP
  {\bf 05} (2015) 132, \href{http://arxiv.org/abs/1412.6087}{{\tt
  arXiv:1412.6087}}.

\bibitem{hayden2007black}
P.~Hayden and J.~Preskill, {\it Black holes as mirrors: quantum information in
  random subsystems},  Journal of High Energy Physics {\bf 2007} (2007), no.~09
  120.

\bibitem{Sekino2008}
Y.~Sekino and L.~Susskind, {\it {Fast Scramblers}},  JHEP {\bf 10} (2008) 065,
  \href{http://arxiv.org/abs/0808.2096}{{\tt arXiv:0808.2096}}.

\bibitem{Susskind2011}
L.~Susskind, {\it {Addendum to Fast Scramblers}},
  \href{http://arxiv.org/abs/1101.6048}{{\tt arXiv:1101.6048}}.

\bibitem{Lashkari2011}
N.~Lashkari, D.~Stanford, M.~Hastings, T.~Osborne, and P.~Hayden, {\it {Towards
  the Fast Scrambling Conjecture}},  JHEP {\bf 04} (2013) 022,
  \href{http://arxiv.org/abs/1111.6580}{{\tt arXiv:1111.6580}}.

\bibitem{Maldacena2015}
J.~Maldacena, S.~H. Shenker, and D.~Stanford, {\it {A bound on chaos}},  JHEP
  {\bf 08} (2016) 106, \href{http://arxiv.org/abs/1503.01409}{{\tt
  arXiv:1503.01409}}.

\bibitem{fitzpatrick2016quantum}
A.~L. Fitzpatrick and J.~Kaplan, {\it A quantum correction to chaos},  Journal
  of High Energy Physics {\bf 2016} (2016), no.~5 70.

\bibitem{perlmutter2016bounding}
E.~Perlmutter, {\it Bounding the space of holographic cfts with chaos},
  Journal of High Energy Physics {\bf 2016} (2016), no.~10 69.

\bibitem{Stanford2015}
D.~A. {Roberts} and D.~{Stanford}, {\it {Diagnosing Chaos Using Four-Point
  Functions in Two-Dimensional Conformal Field Theory}},  Physical Review
  Letters {\bf 115} (Sept., 2015) 131603,
  \href{http://arxiv.org/abs/1412.5123}{{\tt arXiv:1412.5123}}.

\bibitem{Fitzpatrick2014}
A.~L. Fitzpatrick, J.~Kaplan, and M.~T. Walters, {\it {Universality of
  Long-Distance AdS Physics from the CFT Bootstrap}},  JHEP {\bf 08} (2014)
  145, \href{http://arxiv.org/abs/1403.6829}{{\tt arXiv:1403.6829}}.

\bibitem{Roberts2014}
D.~A. Roberts, D.~Stanford, and L.~Susskind, {\it {Localized shocks}},  JHEP
  {\bf 03} (2015) 051, \href{http://arxiv.org/abs/1409.8180}{{\tt
  arXiv:1409.8180}}.

\bibitem{Liu2018}
C.~Liu and D.~A. Lowe, {\it {Notes on Scrambling in Conformal Field Theory}},
  \href{http://arxiv.org/abs/1808.09886}{{\tt arXiv:1808.09886}}.

\bibitem{DiFrancesco:1997nk}
P.~Di~Francesco, P.~Mathieu, and D.~Senechal, {\em {Conformal Field Theory}}.
\newblock Graduate Texts in Contemporary Physics. Springer-Verlag, New York,
  1997.

\bibitem{zamolodchikov1987conformal}
A.~B. Zamolodchikov, {\it Conformal symmetry in two-dimensional space:
  recursion representation of conformal block},  Theoretical and Mathematical
  Physics {\bf 73} (1987), no.~1 1088--1093.

\bibitem{Swingle2017}
D.~{Chowdhury} and B.~{Swingle}, {\it {Onset of many-body chaos in the O (N )
  model}},  Physical Review D {\bf 96} (Sept., 2017) 065005,
  \href{http://arxiv.org/abs/1703.02545}{{\tt arXiv:1703.02545}}.

\end{thebibliography}\endgroup
\end{document}